\documentclass{article}
\usepackage{spconf,amsmath,graphicx}
\usepackage{amsfonts}
\usepackage{enumitem}
\usepackage{xcolor}
\usepackage{hyperref}

\title{Property Neurons in Self-Supervised Speech Transformers}

\name{Tzu-Quan Lin$^1$, Guan-Ting Lin$^1$, Hung-yi Lee$^1$, Hao Tang$^2$}

\address{
  $^1$Graduate Institute of Communication Engineering, National Taiwan University, Taiwan\\
  $^2$University of Edinburgh, United Kingdom
}

\begin{document}

\maketitle

\begin{abstract}
There have been many studies on analyzing self-supervised speech Transformers, in particular, with layer-wise analysis.
It is, however, desirable to have an approach that can pinpoint exactly a subset of neurons that is responsible for a particular property of speech, being amenable to model pruning and model editing.
In this work, we identify a set of property neurons in the feedforward layers of Transformers to study how speech-related properties, such as phones, gender, and pitch, are stored.
When removing neurons of a particular property (a simple form of model editing), the respective downstream performance significantly degrades, showing the importance of the property neurons.
We apply this approach to pruning the feedforward layers in Transformers, where most of the model parameters are.
We show that protecting property neurons during pruning is significantly more effective than norm-based pruning.
The code for identifying property neurons is available at \url{https://github.com/nervjack2/PropertyNeurons}.
\end{abstract}

\begin{keywords}
speech self-supervised models, Transformer, neuron analysis
\end{keywords}

\section{Introduction}
Despite the strong performance of self-supervised speech Transformers~\cite{baevski2020wav2vec, hsu2021hubert, chen2022wavlm, lin2023melhubert} on a slew of benchmarks~\cite{yang2021superb, tsai2022superb, feng2023superb}, we happen to know very little about their inner working.
Prior work has largely focused on probing, measuring how accessible phonetic~\cite{pasad2021layer, pasad2023comparative}, prosodic~\cite{lin2023utility}, speaker~\cite{ashihara2024self}, lexical~\cite{pasad2024self} information are.
It is important to know at which layer a particular type of information is the most prominent.
However, these analyses only make use of the fact that these models have multiple layers, limiting of what we can understand if none of the other structures are taken into account.

\begin{figure}
    \centering
    \includegraphics[width=7.5cm]{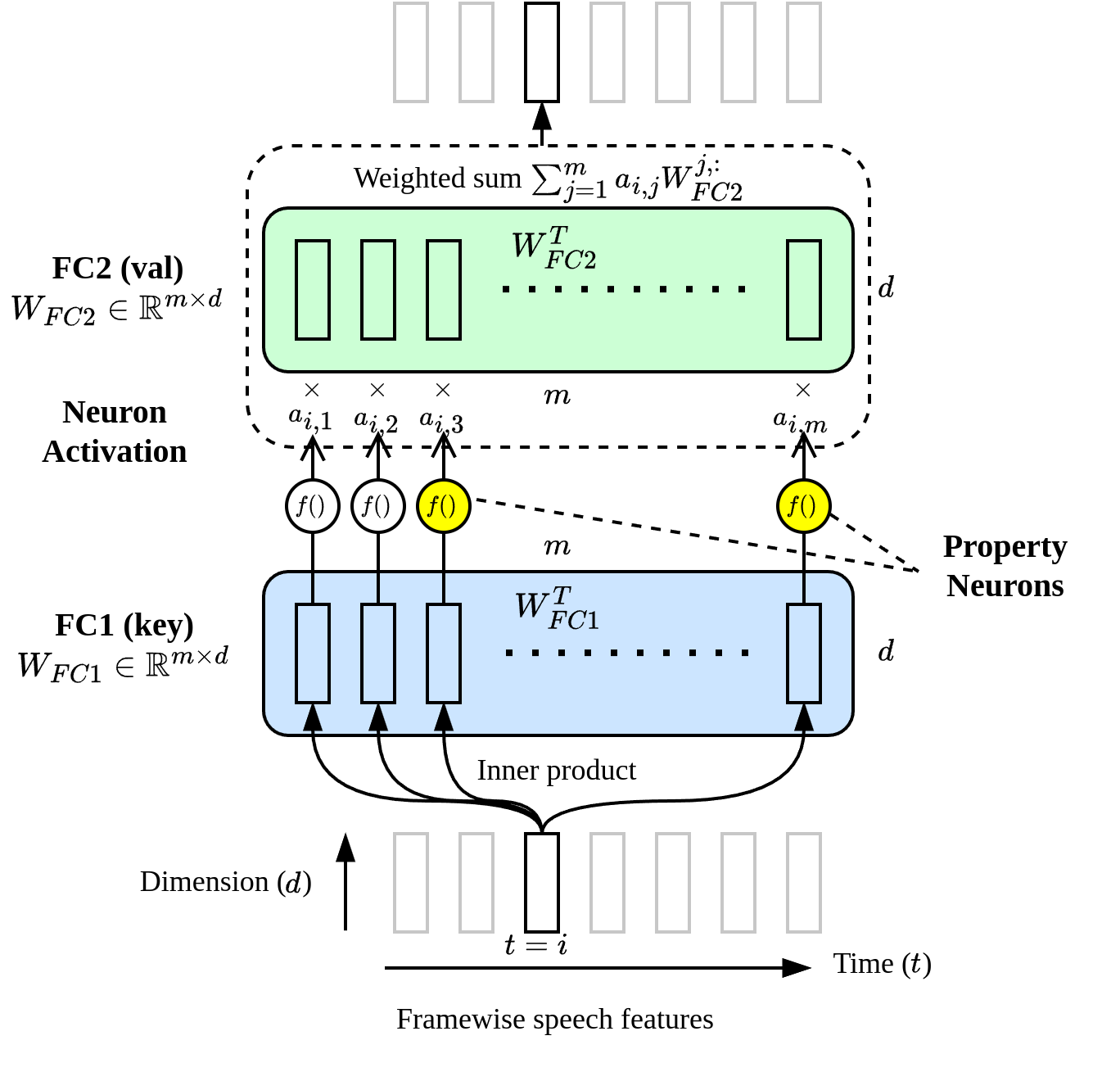}
    \vspace{-1em}
    \caption{The illustration of how feed-forward networks in Transformers could be regard as a type of neural memory.}
    \label{fig:ffn-neuron}
\end{figure}

In this work, we study the simplest structure of these intermediate layers---their coordinates, or more commonly referred to as neurons.
Analyzing neurons is perhaps one of the earliest method for analyzing neural networks (e.g., as used in \cite{hinton1992how}).
When the networks are small, one can visualize the learned filters to study individual neurons \cite{hinton1992how,coates2012learning}.
Other than visualizing the filters, one can also identify the input that leads to high response for a particular neuron~\cite{olah2017feature}, or more generally, correlate the response of a neuron with properties of the input~\cite{bau2018identifying, dalvi2019one, radford2017learning}.
The recent surge in analyzing neurons in Transformers is due to Geva \textit{et al.}~\cite{geva2020transformer}, who interpret the feed-forward layers as key-value memories. 
Subsequent work derived from this viewpoint identifies neurons related with factual knowledge~\cite{dai2021knowledge}, task-specific skills~\cite{wang2022finding}, positional information~\cite{voita2023neurons}, specific languages~\cite{tang2024language}, and privacy information~\cite{chen2024learnable}.
Yet, little, if any, neuron analysis has been done in speech models.

Inspired by the key-value perspective in the feedforward layers of Transformers, in this work, we study properties unique to speech and identify neurons in the feedforward layers that correlate well with phones, gender, and pitch in self-supervised speech Transformers.
For a particular property (such as the one related to phones), we define several groups (e.g., vowels, voiced consonants, and unvoiced consonants).
We then compute the probability of each neuron co-occurring with a phone, and filter out the ones whose probability is lower than a baseline.
In other words, we have a set of neurons that activates when a phone is present in the input (with the definition of being activated to be defined in later sections).
We identify neurons that are specific for each group
and are not activated by phones from other groups.
We refer to these as group neurons.
Finally, we take the union of group neurons from different groups to form a set of property neurons.
In other words, the variation for a particular property is summarized within the discovered set of neurons.

Identifying property neurons has immediate applications, offering opportunities for model editing and model pruning.
As an example, we find that our model fails to identify female speakers after clamping (a simple form of model editing) the group neurons associated with female, while having minimal impact on identifying male speakers.
This shows that the neurons are indeed important for identifying female speakers.
As another example, we can improve model compression by protecting property neurons during model pruning. 
In sum, in addition to the insights it provides, our proposed analysis has applications to model editing and model pruning that are not possible with layer-wise probing.
\section{Feedforward layers of Transformers}

A Transformer consists of multiple Transformer blocks \cite{vaswani2017attention}, each of which has two feedforward layers.
A layer, for example in layer-wise studies \cite{pasad2021layer}, usually refers to the output of the second feedforward layer.
The dimension of the hidden vector between the two feedforward layers are typically much larger the rest of the model, so the two feedforward layers take up most of the parameters of a Transformer.
All in all, the two feedforward layers play an important role in Transformers, and deserve more attention than they already have.

Geva \textit{et al.}~\cite{geva2020transformer} propose to view feedforward layers in Transformers as key-value memories. 
The concept of neural memory \cite{sukhbaatar2015end} is widely used in deep learning.
Self-attention in Transformer blocks is an example \cite{vaswani2017attention}.
Neural memory is composed of $m$ key-value pairs $(k_1, v_1), \dots, (k_m, v_m)$, where each key $k_i$ and each value $v_i$ are $d$-dimensional vectors.
The keys and values can be stacked row-wise to form matrices $K \in \mathbb{R}^{m\times d}$ and $V \in \mathbb{R}^{m\times d}$ 
Given an input query $q \in \mathbb{R}^{d}$, we calculate the distribution, $\text{softmax}(q K^\top)$, across the keys $K$, and use it to compute a weighted sum over the values
\begin{align}
\text{attn}(q) = \text{softmax}(q K^\top) V.
\end{align}
In neural memory, keys are responsible for capturing input patterns, whereas values serve as slots for storing memories.

The computation of the feedforward layers, on the other hand, is
\begin{align}
\text{FFN}(x) = f(x W_{\text{FC1}}^\top) W_{FC2}
\end{align}
where $W_{FC1} \in \mathbb{R}^{m \times d}$ and $W_{FC2} \in \mathbb{R}^{m \times d}$ denote the weight matrix of the first and second feed-forward layers, and $f$ is the activation function.
Geva \textit{et al.}~\cite{geva2020transformer} argue that the computation of the feeforward layers fit the perspective of key-value memories.
Following this view, we will study the output of the \emph{first} feedforward layer rather than the second.
We are particularly interested in the coordinates, because they correspond to how much a key is activated or matched.
\section{Neuron activations}
\label{sec: neurons-capture-property}

To analyze when the neurons activate, we need to define what it means for a neuron to be activated and how neuron activations correlate with aspects of the input speech.
Below as we introduce the definitions, we will also provide preliminary experiments.
We will use the MelHuBERT \cite{lin2023melhubert} pretrained on 960 hours of LibriSpeech \cite{panayotov2015librispeech}.
All the preliminary results will be on the dev-clean subset of Librispeech.

\vspace{-0.35em}
\subsection{A definition of activation}

For neurons that uses sigmoid or ReLU as activation functions, there is a clear and intuitive definition of when a neuron is activated.
However, it is not so clear for more general activations, such as GELU \cite{hendrycks2016gaussian}.
We instead opt for a ranking approach.
Since the focus is to analyze the first feedforward layer, we refer to $|f(x W_{\text{FC1}}^\top)|$ as the activation values of the first feedforward layer.

\begin{figure}
  \centering
  \includegraphics[width=7cm]{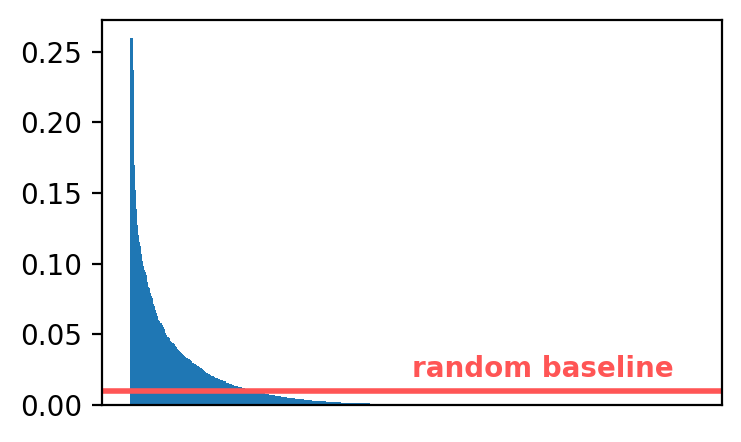}
  \vspace{-0.5em}
  \caption{The probability of neurons activated when a phone [ah] is present.
  The neurons are sorted according to the probability.
  \label{fig:match_prob}}
\end{figure}

We iterate over utterances paired with forced alignments.
For every frame that is labeled with phone $k$, if a neuron (dimension) $i$ is ranked top $\lambda\%$ (here we use $\lambda=1$) based on the activation values of that frame, we say that the neuron $i$ activates when the phone $k$ is present.
After iterating over the set of utterances, we can compute how often a neuron is activated when a phone is present.
In Figure \ref{fig:match_prob}, we show the probability of neurons being activated when the phone [ah] is present.
It is clear that some neurons get activated more frequently than others when [ah] is present. 

\vspace{-0.35em}
\subsection{Activation patterns of properties}
\label{sec: property-group}

In the previous section, we identify neurons that are activated when a particular phone is present.
Typically, only a small set of neurons is activated for each phone (the ones that have higher probability than chance in Figure~\ref{fig:match_prob}).
Formally, for a phone $k$, $S_k$ consists of neuron $j$ such that
\begin{align}
    p(\text{neuron $j$ is activated} \mid \text{the frame is labeled $k$}) > \lambda\%.
\end{align}
In other words, $S_k$ is the set of neurons whose probability of being activated when phone $k$ is present is higher than $\lambda\%$.  
For a phone set $\mathcal{P}$,
the set $S = \bigcup_{p \in \mathcal{P}} S_p$ is the set of the neurons that are involved in identifying phones in the input speech.
Note that $|S|$ might be smaller than the total number of neurons (the total number of dimensions), because not all neurons are involved in identifying phones.
Within the set of neurons $S$, we know that $S_k$ are the ones responsible for identifying phone $k$.
We can represent $S_k$ as a binary vector $v_k \in \{0, 1\}^{|S|}$, where $(v_k)_i = 1$ if neuron $i \in S_k$.
We refer to the binary vector $v_k$ as the \textbf{activation pattern} of phone $k$.

Activation patterns can be conditioned, and we simply add more condition to the probability when finding the activation patterns.
For example, the activation pattern of phone [ah] by a female speaker consists of neuron $j$ such that
$p(\text{neuron $j$ is activated} \mid \text{the frame is labeled [ah]}, \allowbreak \text{the speaker is female})$ is over $\lambda \%$.
Overall, we compute activation patterns of phones conditioned on broad phone classes, gender, and pitch as follows. 

\vspace{0.5em}
\noindent\textbf{Phone classes} \hspace{0.2em}
We group phones into vowels, voiced consonants, and unvoiced consonants, each of which has 15, 15, and 9 phones, respectively. 
We follow ARPABET\footnote{http://www.speech.cs.cmu.edu/cgi-bin/cmudict} and discard lexical stress.
Semi-vowels, such as [r], [y], [w], and [l], are categorized as voiced consonants here, but regardless, in the results we are about to show, they lie in the middle between vowels and consonants. 

\vspace{0.5em}
\noindent\textbf{Gender} \hspace{0.2em}
We compute the activation patterns of phones conditioned on the gender of the speaker. 

\vspace{0.5em}
\noindent\textbf{Pitch} \hspace{0.2em}
We iterate over a set of utterances and divide the pitch range based on the tertiles into ones less than 129.03 Hz, ones between 129.03 and 179.78 Hz, and ones greater than 179.78 Hz.
We compute activation patterns of phones (excluding the unvoiced consonants) conditioned on one of the pitch ranges.

\begin{figure*}
    \centering
    \begin{tabular}{ccc}
         \includegraphics[width=5.5cm]{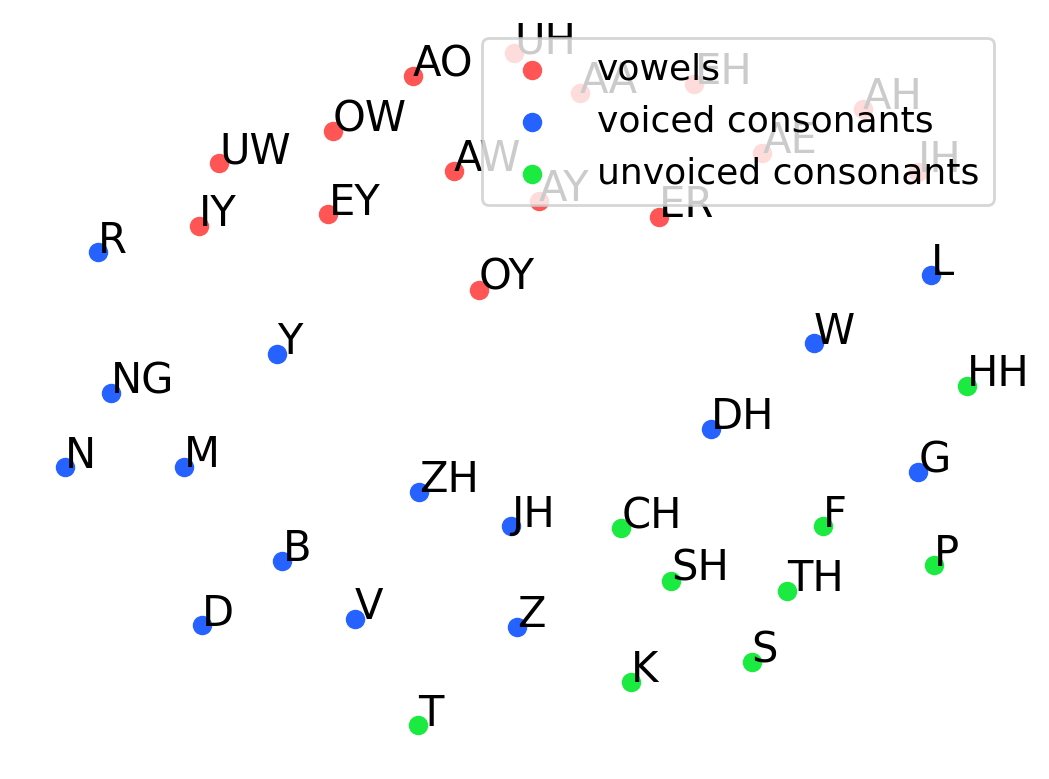} &
        \includegraphics[width=5.5cm]{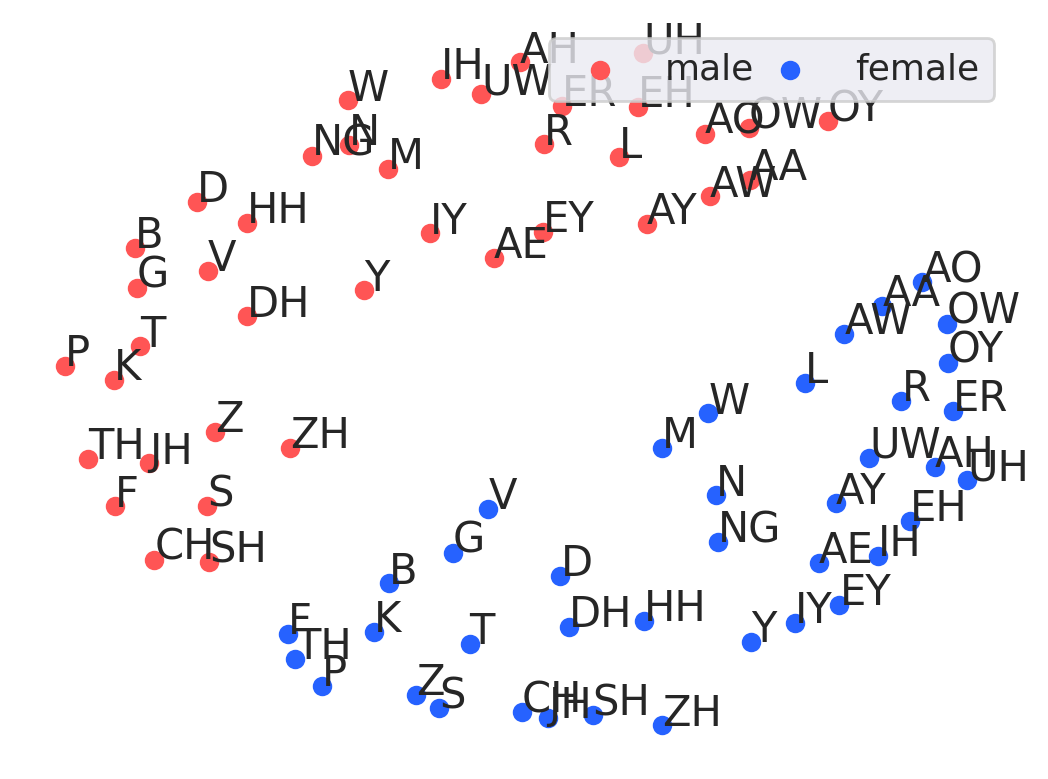} &
        \includegraphics[width=5.5cm]{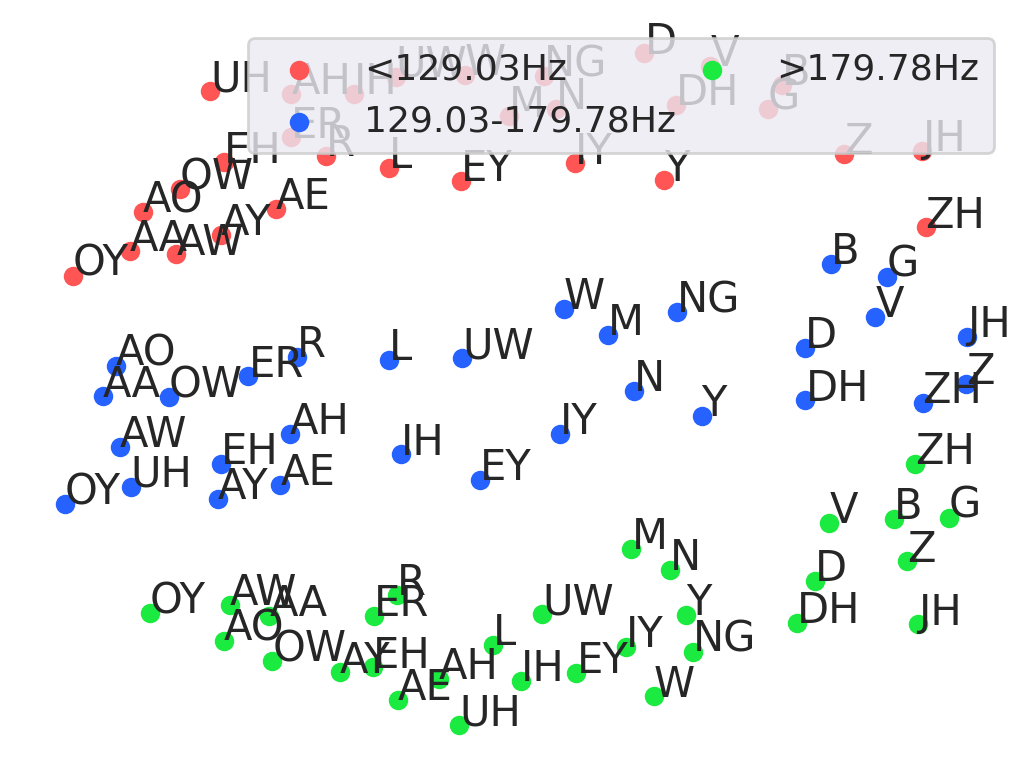} \\
        (a) Phone classes &
        (b) Gender &
        (c) Pitch 
    \end{tabular}
    \caption{The results of multidimensional scaling on the activation patterns of phones conditioned on broad phone classes, gender and pitch.
    Different colors represent different groups.
    For each condition, we show the layer with the highest silhouette score \cite{rousseeuw1987silhouettes}, i.e., the 8th layer, the 1st layer, and the 1st layer, respectively. We consider [r], [y], [w] and [l] as voiced consonants here.}
    \label{fig:multidimensional-scaling}
\end{figure*}

\vspace{0.5em}
For each condition, we apply multidimensional scaling (MDS) \cite{kruskal1978multidimensional} to the activation patterns, and the results are shown in Figure \ref{fig:multidimensional-scaling}.
We find that the activation patterns of phones not only preserves the conditions well, but also respects the similarity among phones.
For example, semi-vowels are placed in the middle between vowels and consonants; nasals are grouped together; diphthongs are grouped together.
Since there is clear cluster structure in the low-dimensional space after MDS, we can use the silhouette score \cite{rousseeuw1987silhouettes} as a measure of cluster tightness; the higher it is, the tighter the cluster.

\vspace{-0.3em}
\subsection{Layer-wise analysis of activation patterns}

\begin{figure}
    \centering
    \includegraphics[width=7.5cm]{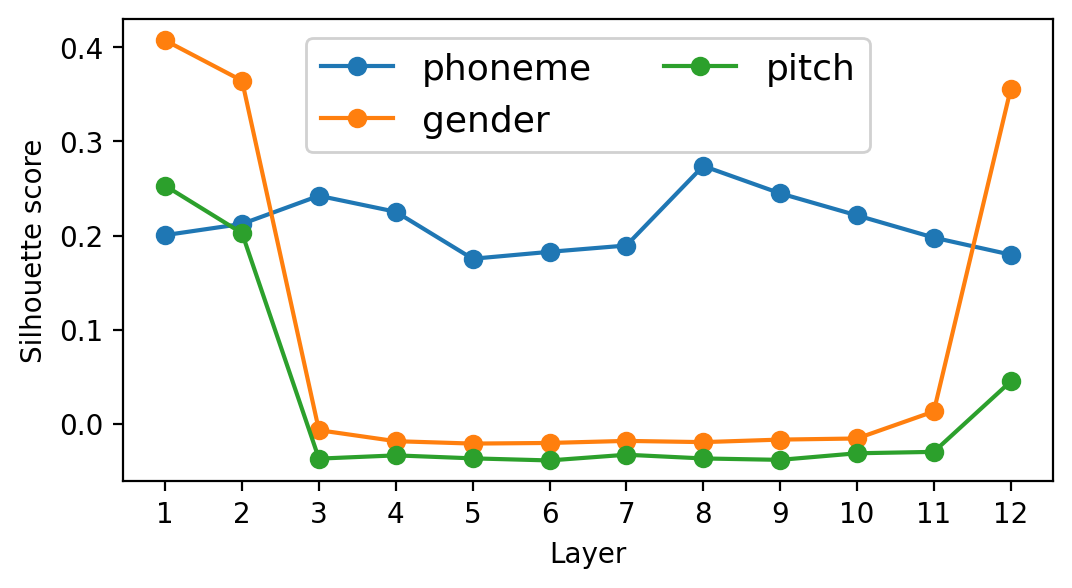}
    \caption{The result of performing multidimensional scaling on the activation patterns of phones for different properties of speech. We report silhouette score to measure cluster tightness.}
    \label{fig:layer-compare}
\end{figure}

We have seen that activation patterns of phones is a useful tool for studying properties of speech.
We can apply the same analysis to different layers of MelHuBERT.
The result is shown in Figure \ref{fig:layer-compare}. 
For phones, all layers exhibit good clustering results, with the 8th layers being the highest.
For gender and pitch, tight clustering results are found in the first two layers and the last layer.
Our results are consistent with other layer-wise analysis \cite{hsu2021hubert, lin2023melhubert, pasad2021layer, lin2023utility, ashihara2024self}.
The neuron analysis presented here can be seen as another form to probing, but without the hassle of training classifiers.
Our approach can also provide the exact activation pattern when a phone is present, not possible to achieve with probing classifiers.

\vspace{-0.4em}
\subsection{Layer-wise analysis of other speech models}

To show the generality of our approach, we examine MelHuBERT \cite{lin2023melhubert}, HuBERT \cite{hsu2021hubert}, wav2vec 2.0 \cite{baevski2020wav2vec}, and WavLM \cite{chen2022wavlm}.
In fact, our approach is not restricted to self-supervised models and can be applied to supervised models as well.
To showcase, we fine-tune MelHuBERT on Librispeech 100 hours subset for phoneme recognition (PR) and Voxceleb1 \cite{nagrani2020voxceleb} for speaker identification (SID).

\begin{figure}
    \centering
    \includegraphics[width=8.0cm]{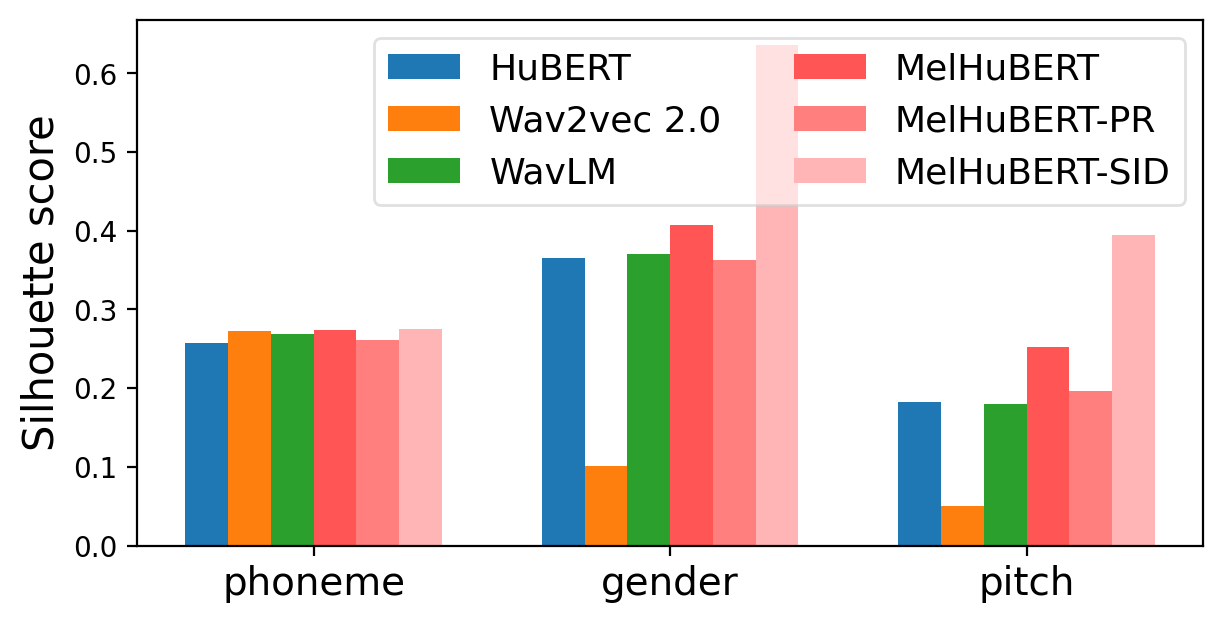}
    \caption{The silhouette score of multidimensional scaling on the activation patterns of phones for different speech models. We report the highest score among all layers for each model and each property. MelHuBERT-PR and MelHuBERT-SID denote fine-tuned MelHuBERT on phoneme recognition and speaker identification respectively.}
     \label{fig:models-compare}
\end{figure}

For different properties of speech and models, we only report the best silhouette score among all layers. 
The experiment results are shown in Figure \ref{fig:models-compare}.
In general, we can discover neurons that identify phones, gender, and pitch in most models.
The only exception is wav2vec 2.0, which scores particularly low in gender and pitch.
For fine-tuned models, we find that fine-tuning on phone recognition does not show a significant improvement in the scores for phones, suggesting that the activation patterns of phones does not change much before and after fine-tuning.
Fine-tuning on speaker identification, however, significantly changes the activation patterns, making clusters in gender and pitch a lot tighter than that before fine-tuning.

\vspace{-0.15em}
\section{Property Neurons}

In Section \ref{sec: neurons-capture-property}, we show that the activation patterns reveal various properties of the input speech. 
In this section, we further identify specific neurons that are particularly important for a specific property, which we refer to as \textbf{property neurons}. In this section, we will again use MelHuBERT as an example, but the approach is applicable to other Transformer-based speech models.

\vspace{-0.3em}
\subsection{Finding Property Neurons}
As described in Section \ref{sec: property-group}, given a specific property of speech (e.g., gender), we can define several groups (e.g., male and female).
For each group, we compute activation patterns for each phone in the group. 
Then, for the $i$-th group, we identify a set of neurons $N_i$ that are activated by a sufficient number of phones (in our case 80\%) in the group.
Next, we obtain the group neurons $G_i$ by computing the difference
\begin{align}
\label{eq: group-neurons}
    G_i = N_i \mathbin{\Big\backslash} \bigcup_{\substack{j=1 \\ j \neq i}}^{n} N_j
\end{align}
where $n$ represents the number of groups for the property.
In words, $G_i$ is the set of neurons that are activated specifically by the $i$-th group and not by any other groups.
Finally, we can obtain the property neurons $P$ by calculating the union of each group neurons
\begin{align}
\label{eq: property-neurons}
    P = \bigcup_{i=1}^{n} G_i
\end{align}
Note that phones, gender, and pitch have their own property neurons.
Neurons for a particular property is typically a small subset of all the neurons (dimensions).

\vspace{-0.3em}
\subsection{Do property neurons really encode property?}
\label{sec: neurons-encode-property}

We verify whether property neurons identified this way actually encode the information of properties for both self-supervised and fine-tuned models.
Pruning the feedforward layers not only can tell the importance of the discovered neurons, but is also practical for other applications that have memory or computation constraints \cite{lin2022compressing}.

\begin{figure}
    \centering
    \includegraphics[width=8.5cm]{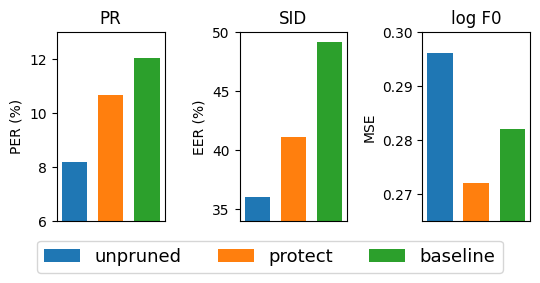}
    \vspace{-0.5em}
    \caption{The result of protecting property neurons of phones, gender, and pitch for MelHuBERT during task-agnostic pruning. We fine-tune the models with self-supervised pre-training objective until converge after pruning.}
     \label{fig:pruning-bar}
\end{figure}

For self-supervised models, we prune the feedforward layers together in the entire model.
First, we calculate property neurons for phones, gender, and pitch for MelHuBERT with Equation \ref{eq: property-neurons} on the 100-hour subset of Librispeech.
For phones, we consider both grouping by broad phone classes as described in Section \ref{sec: property-group} and treating individual phones as their own group.
For all layers, we prune neurons other than the property neurons of phones, gender, and pitch. 
Finally, we fine-tune the model on the full Librispeech 960 hours until convergence with the self-supervised pre-training objective as is done for regular model pruning.
Following \cite{lin2022compressing}, 
when pruning the $i$-th neuron, 
we prune the $i$-th column of the first feedforward
layer $W_{FC1}^i$ and the $i$-th row of the second feedforward
layer $W_{FC2}^i$.
As a baseline, we use the L1 norm of the weights magnitude $\| W_{FC1}^i\|_1 + \|W_{FC2}^i\|_1$ as a criterion, pruning neurons with the smallest L1 norm.
For both approaches, we prune about 80\% of the neurons in the model.
The result is shown in Figure \ref{fig:pruning-bar}.
It can be seen that compared to the baseline pruning method, protecting property neurons significantly reduces performance loss during the pruning process. The results are consistent in PR, SID, and f0 reconstruction.

For supervised models, we examine models fine-tuned on Voxceleb1 for speaker identification.
We compute the group neurons related to male and female in the fine-tuned model with Equation \ref{eq: group-neurons}. 
We replace the columns in $W_{FC2}^{i}$ (also referred to as values in Geva \textit{et al.}~\cite{geva2020transformer}) corresponding to the group neurons of either male or female with zero vectors.
The result is shown in Table \ref{tab:erase-gender}. 
It can be seen that after ``erasing'' the values related to female, 
the identification error rate for female increases significantly compared to male. 
Conversely, erasing the values related to male results in a substantial increase in the error rate for male. 
This indicates that the group neurons we have identified information specific to that group.

\begin{table}[h]
    \centering
    \begin{tabular}{|c|c|c|}
        \hline
         & Male ($\bigtriangleup$ERR) & Female ($\bigtriangleup$ERR) \\ \hline
        Erase Male & 22.43 & 2.24  \\ \hline
        Erase Female & 4.1 & 18.58  \\ \hline
    \end{tabular}
    \vspace{0.1em}
    \caption{The changes in the identification error rates after erasing the values slots of male or female's group neurons in a supervised fine-tuned speaker identification model.}
    \label{tab:erase-gender}
\end{table}

\begin{figure}[t]
    \centering
    \includegraphics[width=7.5cm]{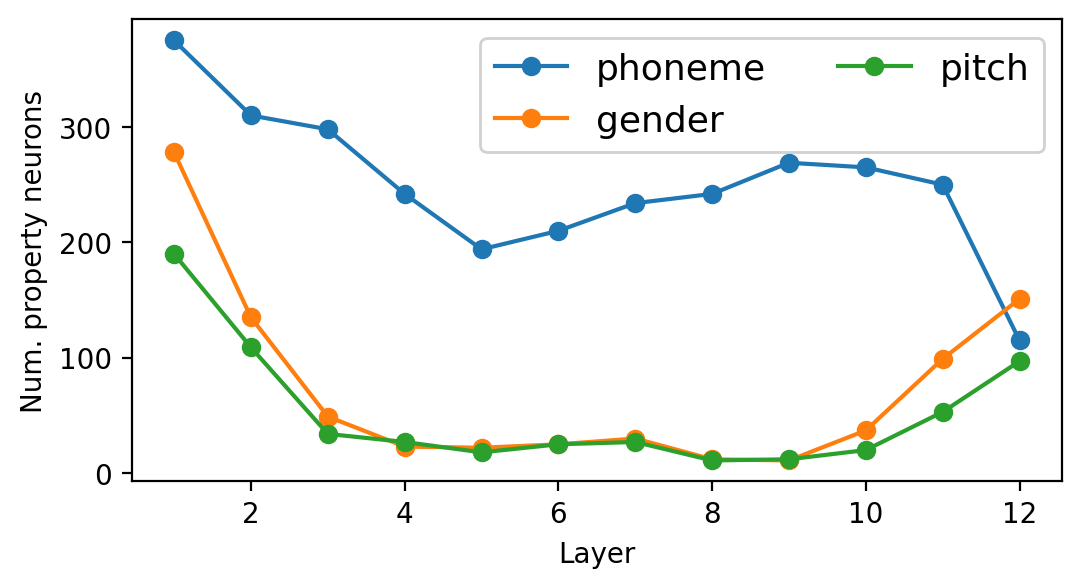}
    \vspace{-0.5em}
    \caption{The number of property neurons in different layers of MelHuBERT.}
     \label{fig:num-property-neurons}
\end{figure}

\vspace{-0.3em}
\subsection{How many property neurons are there?}

From the layer-wise analysis, we know that information is processed differently at different layers, and we are interested in how the number of property neurons changes over the layers.
If they change, whether they show any consistent trend.
Given the pruning results before, these numbers can inform us how much pruning is possible on the individual layers.
The result on Librispeech dev-clean subset is shown in Figure \ref{fig:num-property-neurons}. 

First, it is evident that the number of property neurons is largely related to the ability of recognizing the property. 
For example, as shown in Figure \ref{fig:layer-compare}, the middle layers have a low silhouette score for properties like gender and pitch, and the number of property neurons for gender and pitch in these layers are significant fewer as well. 
Additionally, we find that compare to the last layer, the earlier layers require a significantly larger number of neurons for identifying properties.
This might be related to the accessibility of the information at each layer.
Phones, gender, and pitch information are harder to access in early layers. 

\vspace{-0.3em}
\subsection{Some neurons encode more than one property}

\begin{figure}
    \centering
    \includegraphics[width=5.5cm]{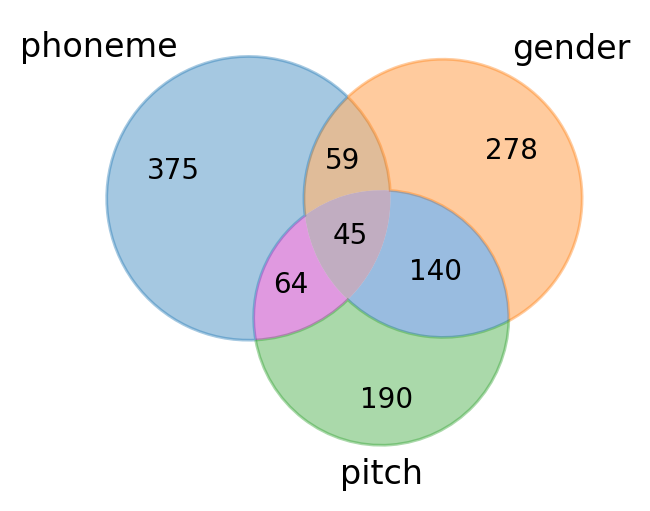}
    \vspace{-0.5em}
    \caption{The number of property neurons for different properties in the first layer of MelHuBERT.}
     \label{fig:venn-overlap}
\end{figure}

Given that different properties inherently correlate with each other, we are interested in how much overlap there is among the property neurons for different properties.
The results on Librispeech dev-clean subset for the first layer of MelHuBERT are shown in Figure \ref{fig:venn-overlap}.
The observation of other layers are similar to the first. 
There is indeed some overlapping between different properties. 
The extent of overlapping varies among properties. 
For example, gender and pitch have a higher number of overlapping property neurons, a reasonable result given how correlated the two properties are, i.e., knowing the gender gives information about the average pitch and vice versa.
Moreover, it can be seen that the union of the property neurons for phones, gender, and pitch is much smaller than the total number of neurons in the feed-forward networks (3072).
The property neurons not in the these sets could potentially be pruned, consistent with the model pruning results before.

\section{Application of Property Neurons}

The biggest strength of our approach is that we can pinpoint exactly the set of neurons for a particular property of speech, amenable to applications such as model editing.
Below we present two example applications.

\vspace{-0.3em}
\subsection{Improving task-specific pruning}

\begin{figure}
    \centering
    \includegraphics[width=7.5cm]{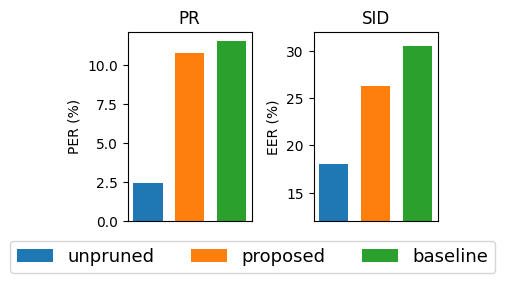}
    \vspace{-0.5em}
    \caption{The results of protecting property neurons during task-specific pruning. We protect property neurons of phones during PR and gender during SID.}
     \label{fig:task-specific-pruning}
\end{figure}

A simple application of property neurons is to enhance the performance of supervised models during task-specific pruning.
We use the same pruning method similar in Section \ref{sec: neurons-encode-property} to prune a fine-tuned model (as opposed to a self-supervised model  in Section \ref{sec: neurons-encode-property}).
We fine-tune MelHuBERT on the 100-hour subset of LibriSpeech for phone recognition (PR) and on Voxceleb1 for speaker identification (SID).
Pruning on the fine-tuned models is done with the property neurons protected. 
For PR, we protect property neurons related to phones, and
for SID, we protect property neurons related to gender.
For phones, we consider both grouping broad phone classes in Section \ref{sec: property-group} and leaving each phone as its own group. 
For both PR and SID, the property neurons are computed on the Librispeech 100-hour subset.
We iteratively prune and fine-tune the model until about 5\% of the neurons remain.
The results are shown in Figure \ref{fig:task-specific-pruning}.  
It can be seen that protecting property neurons during pruning does improve the model performance during task-specific pruning.

\vspace{-0.3em}
\subsection{Erase speaker information for privacy}

As demonstrated in Section \ref{sec: neurons-encode-property}, erasing group neurons associated with a specific group (e.g., male or female) can significantly increase the speaker identification error rate for that group, while the error rate for the other group changes minimally. 
By identifying the neurons associated with a specific speaker, there is potential to erase specific speaker's information from the model without affecting other speakers' performance. This approach could become applicable to research concerning speaker privacy. We regard these possibilities as future work. 
\section{Related Work}

In the speech domain, 
many studies have analyzed the layer-wise features of speech SSL models. 
Pasad \textit{et al.} \cite{pasad2021layer} calculated the similarity between mean-pooled phone-level representations and phone labels. 
Lin \textit{et al.} \cite{lin2023utility} examined the contribution of different layers features to prosody downstream tasks.
Ashihara \textit{et al.} \cite{ashihara2024self} used similar method to analyze the layer-wise distribution of speaker information in speech models.
Compared to prior work, 
they can only identify whether a specific information is present or not in a layer of a model.
Our approach can precisely identify neurons that are responsible for specific properties of speech, 
and our analysis enables applications that were not possible with previous analyses.

In NLP, 
many have studied if and how certain properties are stored within Transformers, and many have followed the approach proposed in Geva \textit{et al.}~\cite{geva2020transformer}.
Dai \textit{et al.} \cite{dai2021knowledge} identified knowledge neurons that store factual knowledge through fill-in-the-blank cloze tasks. 
Wang \textit{et al.} \cite{wang2022finding} found skill neurons that store specific task skills through prompt tuning. 
Voita \textit{et al.} \cite{voita2023neurons} showed that some neurons encode positional information.
Tang \textit{et al.} \cite{tang2024language} identified language-specific neurons by computing activating probability across different languages and neurons. 
Chen \textit{et al.} \cite{chen2024learnable} used learnable binary masks to identify neurons related to personally identifiable information.
In contrast to these studies, we show the utility of the approach once the neurons are identified with model editing and model pruning. 
Additionally, we identify neurons that are particularly important for properties unique to speech.


\section{Conclusion}
In this work, we propose a method to identify property neurons for phones, gender, and pitch.
We present a comprehensive study of the characteristics of property neurons.
When removing the neurons for a particular group, the downstream performance deteriorates, an evidence that the neurons are indeed important for that particular group.
We then show how property neurons can be used for model pruning.
In particular, we protect property neurons in both task-agnostic and task-specific pruning, and we see consistent improvements. 
We believe that property neurons not only serve as a tool for analysis but also provides other opportunities for model editing.

\newpage
\clearpage

\bibliographystyle{IEEEtran}
\bibliography{ref}

\end{document}